\documentclass[sigconf, nonacm]{acmart}

\newcommand\vldbavailabilityurl{https://github.com/UIC-InDeXLab/Needle}
\newcommand\vldbpagestyle{plain}

\usepackage{flushend}
\usepackage[a-1b]{pdfx}   
\usepackage{balance}

\usepackage{color}
\usepackage[labelfont=bf]{caption}
\usepackage{subcaption}
\usepackage{hyperref}
\usepackage{makecell}
\usepackage{fixltx2e}
\usepackage{enumitem}
\usepackage{rotating,multirow}
\usepackage{etoolbox,xspace}
\usepackage{algorithmicx}
\usepackage{algorithm}
\usepackage{fontawesome}
\usepackage{algpseudocode}

\algtext*{EndWhile}
\algtext*{EndIf}
\algtext*{EndFunction}
\algtext*{EndFor}
\algrenewcommand\algorithmicrequire{\textbf{Input:}}
\algrenewcommand\algorithmicensure{\textbf{Output:}}

\usepackage{amsthm}

\usepackage{tabularx}
\usepackage{blindtext}
\usepackage{ragged2e}  
\usepackage{booktabs}  
\usepackage{textcomp}
\usepackage{url}
\usepackage{comment}
\usepackage{enumitem}
\usepackage{nicematrix}
\usepackage[simplified]{pgf-umlcd}
\usepackage{tikz}
\usepackage{ifthen}
\usepackage{diagbox}
\usepackage{listings}

\newcolumntype{Y}{>{\RaggedRight\arraybackslash}X}

\usepackage{booktabs}
\usepackage{xspace}

\newenvironment{proof}{\paragraph{Proof:}}{\hfill$\square$}
\newcounter{example}
\renewcommand{\theexample}{\arabic{example}}

\newcommand{\squishlist}{
 \begin{list}{$\bullet$}
  { \setlength{\itemsep}{0pt}
     \setlength{\parsep}{1pt}
     \setlength{\topsep}{1pt}
     \setlength{\partopsep}{0pt}
     \setlength{\leftmargin}{1em}
     \setlength{\labelwidth}{1em}
     \setlength{\labelsep}{0.5em} } }

\newcommand{\squishend}{
  \end{list}
}

\definecolor{americanrose}{rgb}{1.0, 0.01, 0.24}
\definecolor{airforceblue}{rgb}{0.36, 0.54, 0.66}
\definecolor{ao(english)}{rgb}{0.0, 0.5, 0.0}
\definecolor{ao}{rgb}{0.0, 0.0, 1.0}

\newcommand{\eat}[1]{}

\newcommand{\submit}[1]{}

\usepackage{xspace}

\renewcommand{\marginpar}[2][]{}

\newcommand{\sys}{\textsc{NeedleDB}\xspace}
\newcommand{\cli}{\texttt{needlectl}\xspace}

\usepackage{bbm, dsfont}

\begin{document}

\title{\sys: A Generative-AI Based System for Accurate and Efficient Image Retrieval using Complex Natural Language Queries}

\author{Mahdi Erfanian}
\affiliation{%
  \institution{University of Illinois Chicago}
}
\email{merfan2@uic.edu}

\author{Abolfazl Asudeh}
\affiliation{%
  \institution{University of Illinois Chicago}
}
\email{asudeh@uic.edu}

\begin{abstract}
We demonstrate \sys, an open-source, deployment-ready database system for answering complex natural language queries over image data.
Unlike existing approaches that rely on contrastive-learning embeddings (e.g., CLIP), which degrade on compositional or nuanced queries, \sys leverages generative AI to synthesize \emph{guide images} that represent the query in the visual domain, transforming the text-to-image retrieval problem into a more tractable image-to-image search.
The system aggregates nearest-neighbor results across multiple vision embedders using a weighted rank-fusion strategy grounded in a Monte Carlo estimator with provable error bounds.
\sys ships with a full-featured command-line interface (\cli), a browser-based Web~UI, and a modular microservice architecture backed by PostgreSQL and Milvus.
On challenging benchmarks, it improves Mean Average Precision by up to 93\% over the strongest baseline while maintaining sub-second query latency.
In our demonstration, attendees interact with \sys through three hands-on scenarios that showcase its retrieval capabilities, data ingestion workflow, and pipeline configurability.
\end{abstract}
\maketitle
\pagestyle{\vldbpagestyle}

\ifdefempty{\vldbavailabilityurl}{}{
\vspace{.3cm}
\begingroup\small\noindent\raggedright\textbf{PVLDB Artifact Availability:}\\
The source code, demonstration video, and other artifacts are available at \url{\vldbavailabilityurl}.
\endgroup
}

\vspace{-2mm}
\section{Introduction}

Answering natural language queries over image collections is a fundamental challenge in multi-modal data management.
State-of-the-art approaches use contrastive-learning models such as CLIP~\cite{radford2021learning} and ALIGN~\cite{jia2021scaling} to project both text and images into a shared embedding space and then apply nearest-neighbor search.
While effective for simple object-detection queries (e.g., ``\emph{traffic light}''), these methods break down for complex, compositional queries that involve spatial relationships, attribute descriptions, or abstract concepts~\cite{needle-full}.

\begin{figure}[t]
  \centering
  \includegraphics[width=0.95\columnwidth]{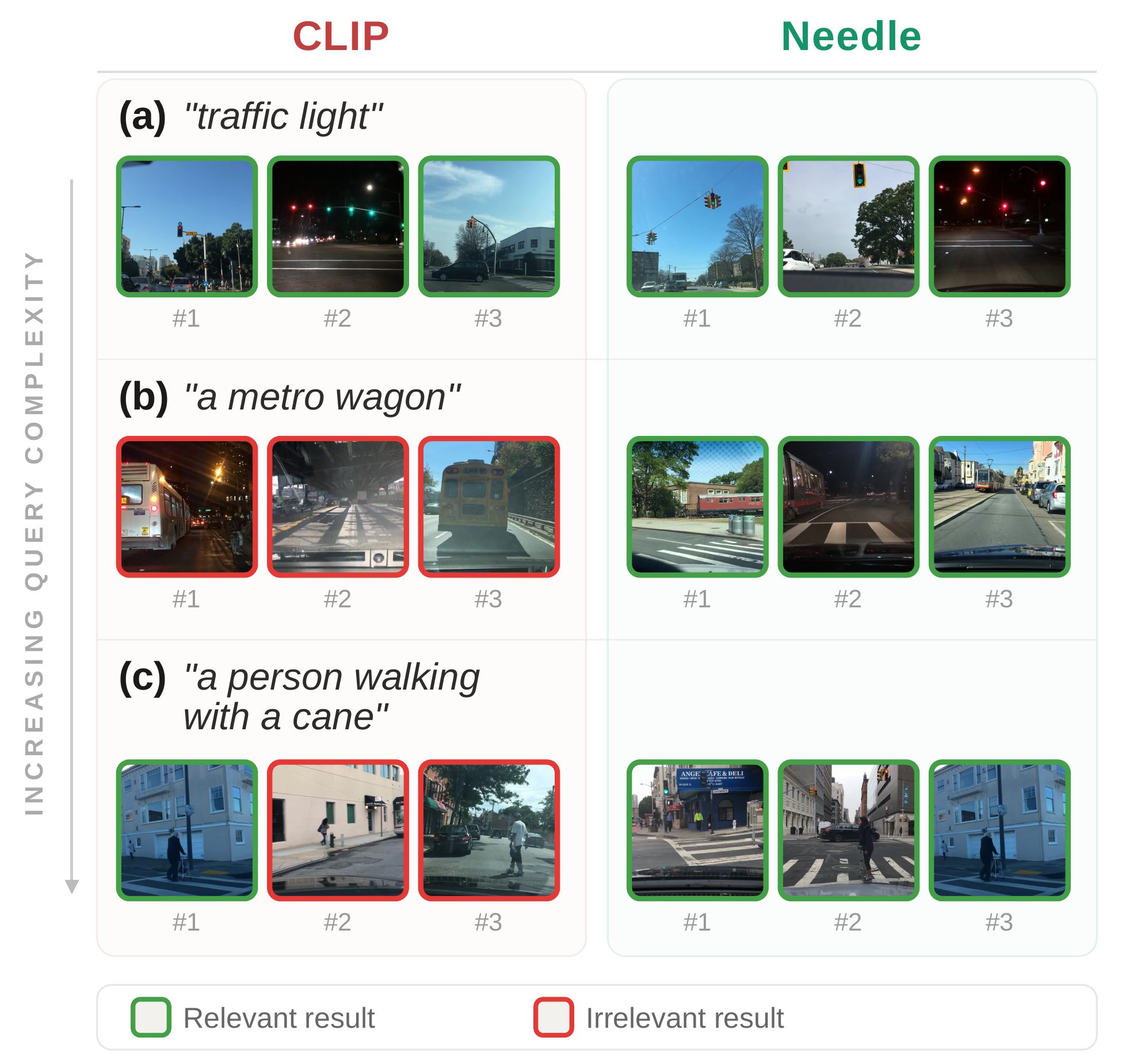}
  \caption{Top-3 results on BDD100K~\cite{yu2020bdd100k} for queries of increasing complexity. CLIP returns relevant results only for the simple query~(a), whereas \sys retrieves relevant images for all three queries using just 2 guide images (RealVisXL) and 2 embedders (EVA, RegNet)
  }
  \label{fig:clip-vs-needle}
  \vspace{-0.5cm}
\end{figure}

Consider the queries of increasing complexity shown in Figure~\ref{fig:clip-vs-needle}: while CLIP retrieves relevant images for a simple query such as \emph{``traffic light''}, it fails entirely on the moderate query \emph{``a metro wagon''} and returns mostly irrelevant results for \emph{``a person walking with a cane''}.
In contrast, \sys retrieves relevant images for all three queries from the BDD100K dataset~\cite{yu2020bdd100k}, using only two guide images generated by RealVisXL and two vision embedders (EVA~\cite{fang2023eva}, RegNet~\cite{xu2022regnet}).
The key insight behind \sys is to reframe text-to-image retrieval as \emph{image-to-image} retrieval: a generative foundation model produces synthetic \emph{guide images} that capture the semantics of the query in the visual domain.
A collection of vision embedders then conducts independent nearest-neighbor searches, and a weighted rank-aggregation step fuses the results.
This procedure is formally grounded in a Monte Carlo estimator whose error decreases exponentially with the number of guide images and embedders~\cite{needle-full}.

In this demonstration we present \sys as a complete, interactive platform. Attendees will (i)~issue their own natural language queries and compare the results to those of CLIP, (ii)~index a live image dataset and search it in real time, and (iii)~explore how the pipeline's hyperparameters---number of guide images, image generation quality, and choice of foundation model---affect retrieval quality.

\vspace{-2mm}
\section{System Architecture}\label{sec:overview}
\begin{figure}[t]
  \centering
  \includegraphics[width=\columnwidth]{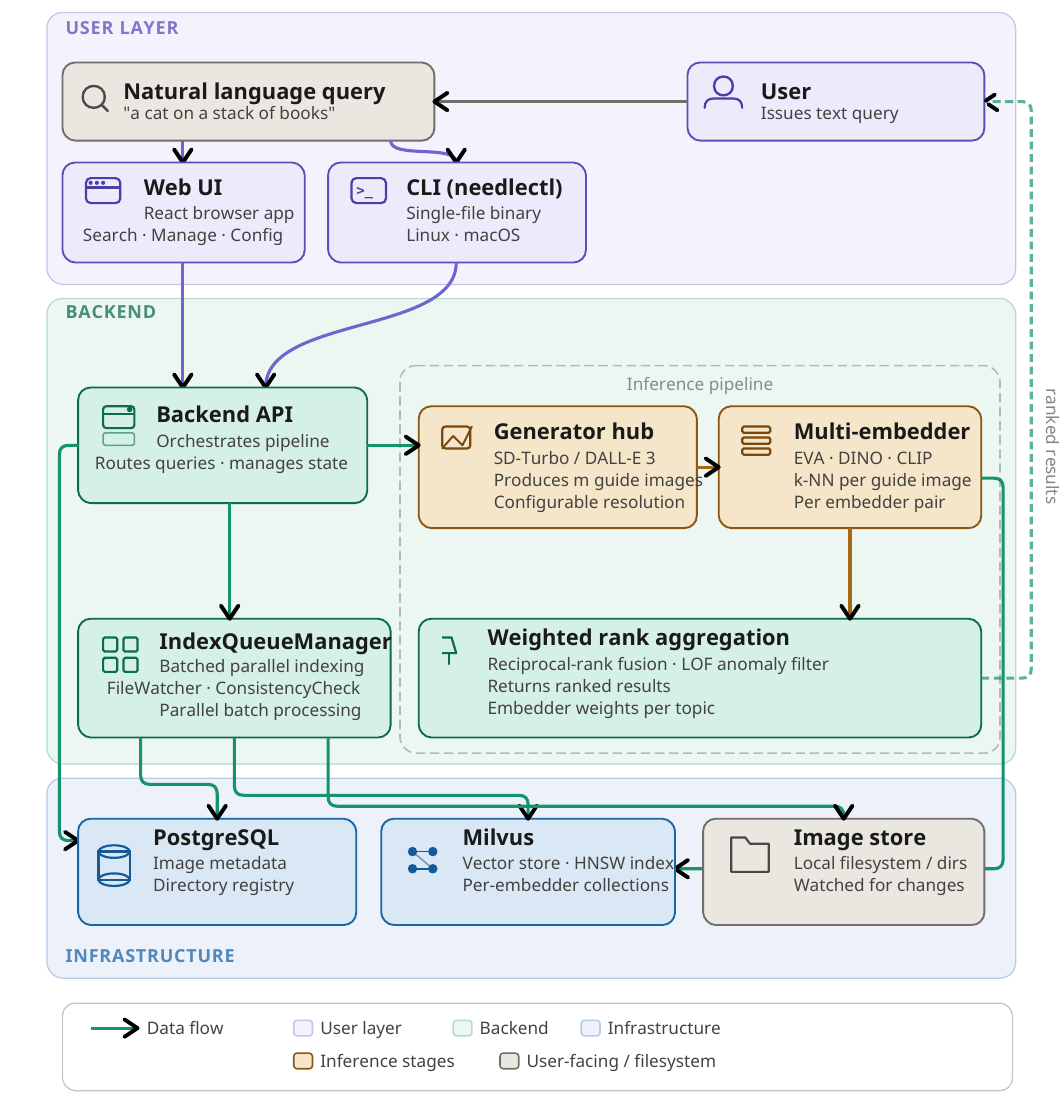}
  \caption{Architecture of the \sys system.  A query flows from the user layer through the backend's multi-stage inference pipeline to the infrastructure services.}
  \label{fig:architecture}
  \vspace{-0.4cm}
\end{figure}

\vspace{-1mm}
\sys follows a microservice architecture with three layers (Figure~\ref{fig:architecture}): a \emph{user-facing layer} (CLI and Web~UI), a \emph{backend API}, and an \emph{infrastructure layer} (vector store, relational database, and image generation service).

\vspace{-2mm}
\subsection{Preprocessing (Indexing)}
\vspace{-1mm}
A user registers one or more image directories via the CLI or Web~UI.
\sys recursively scans each directory, records image metadata in PostgreSQL, and computes embeddings for every image using a configurable ensemble of vision models (e.g., EVA~\cite{fang2023eva}, RegNet~\cite{xu2022regnet}, DINO~\cite{oquab2023dinov2}, CLIP~\cite{radford2021learning}, ConvNeXtV2~\cite{woo2023convnext}, BeViT~\cite{bao2021beit}).
Embeddings are inserted into Milvus~\cite{wang2021milvus} with an HNSW index~\cite{malkov2018efficient} ($M{=}48$, $\mathit{efConstruction}{=}200$) for efficient approximate nearest-neighbor search.
A background file watcher and consistency checker keep the index synchronized with filesystem changes.

\vspace{-3mm}
\subsection{Inference Pipeline}
At query time, \sys processes each query through an optimized two-stage pipeline:

\begin{enumerate}[leftmargin=*, nosep]
  \item \textbf{Guide Image Generation and Multi-Embedder Search.}
    The system invokes the \emph{Generator Hub} service to produce $m$ synthetic guide images using one or more text-to-image foundation models (e.g., SD-Turbo~\cite{sauer2024adversarial}, DALL-E~3~\cite{betker2023improving}).
    By using locally hosted latent-consistency models such as SD-Turbo with 1--4 inference steps at $512{\times}512$ resolution, generation completes in milliseconds rather than seconds.
    Each guide image is then embedded by $l$ vision models selected at installation time, and a $k$-NN search is performed per (guide~image, embedder) pair.
  \item \textbf{Weighted Rank Aggregation.}
    Results from all $(m \times l)$ searches are combined using weighted reciprocal-rank fusion, where each embedder's weight reflects its topic-specific reliability.
    An anomaly detection module based on Local Outlier Factor filters low-quality guide images before aggregation.
\end{enumerate}

\subsection{System Features}

\noindent\textbf{One-Liner Installation.}
\sys can be installed on Linux or macOS with a single command that detects the host OS and GPU (NVIDIA CUDA or Apple MPS), downloads all dependencies, spins up Docker infrastructure, and installs the \cli binary:

{\small
\begin{verbatim}
curl -fsSL https://raw.githubusercontent.com/
  UIC-InDeXLab/Needle/main/scripts/
  install-oneliner.sh | bash -s fast
\end{verbatim}
}

\begin{figure*}[t]
  \centering
  \begin{subfigure}{0.24\textwidth}
    \centering
    \includegraphics[height=1.95cm, keepaspectratio]{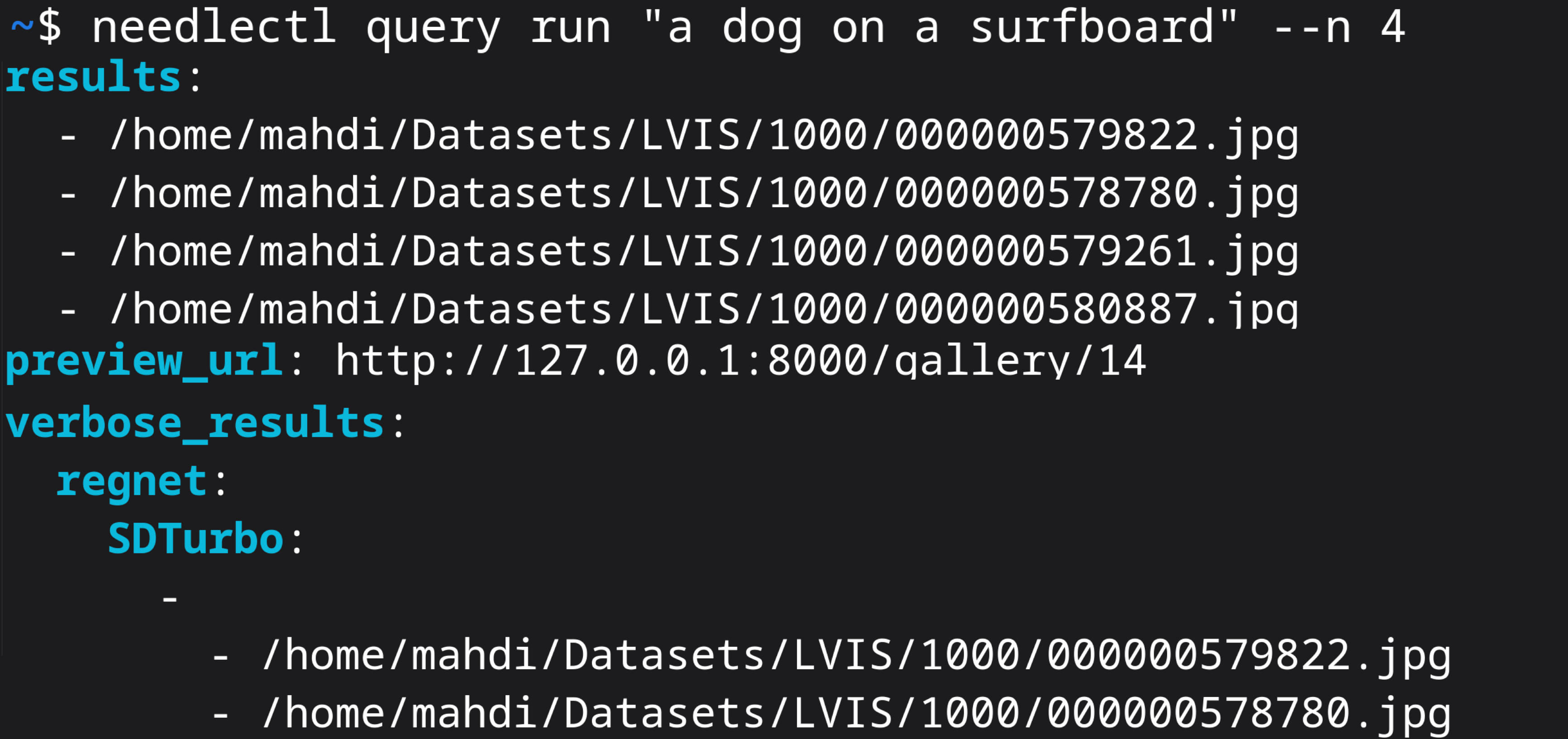}
    \caption{Query execution.}
    \label{fig:cli_search}
  \end{subfigure}
  \hfill
  \begin{subfigure}{0.24\textwidth}
    \centering
    \includegraphics[height=1.95cm, keepaspectratio]{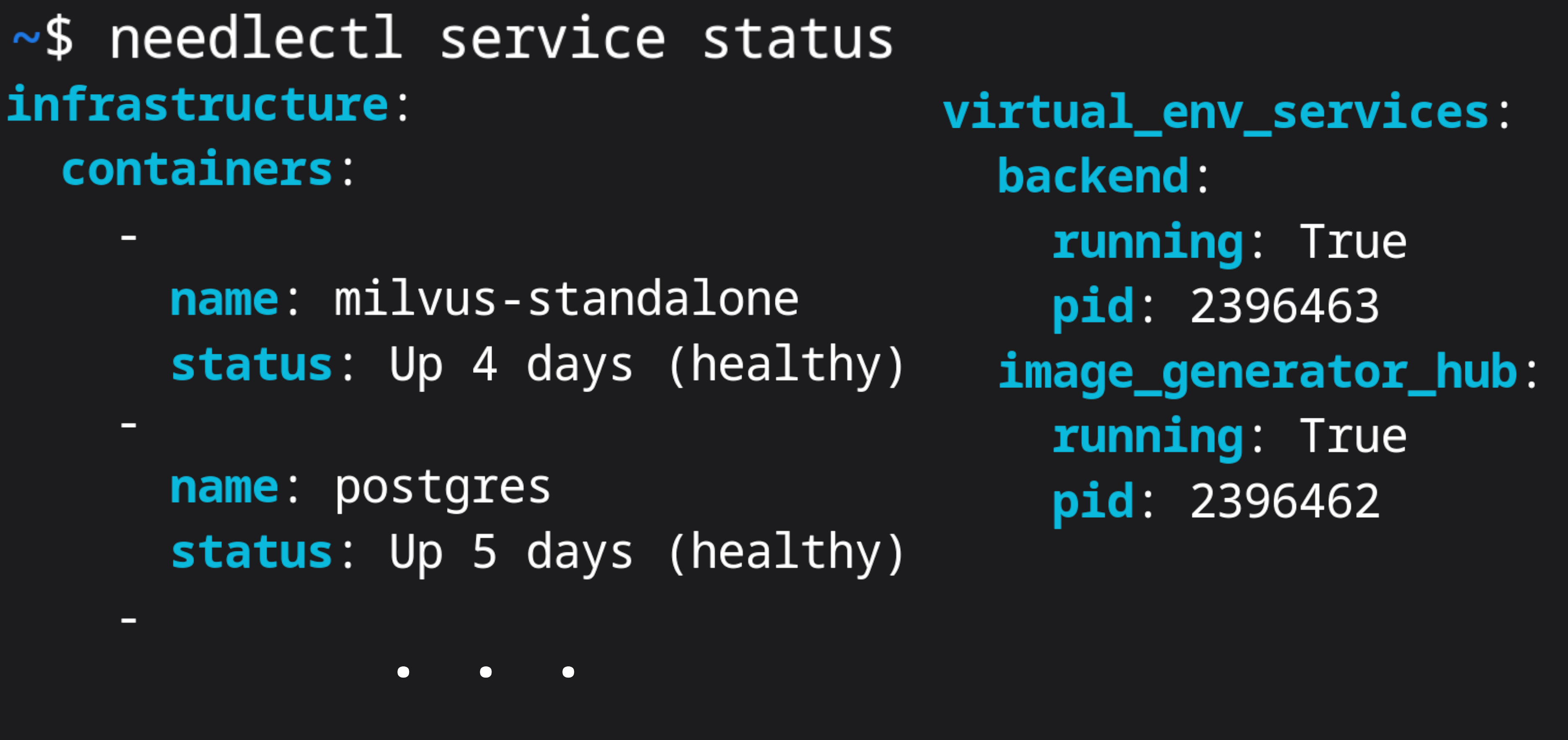}
    \caption{Service status.}
    \label{fig:cli_status}
  \end{subfigure}
  \hfill
  \begin{subfigure}{0.24\textwidth}
    \centering
    \includegraphics[height=1.95cm, keepaspectratio]{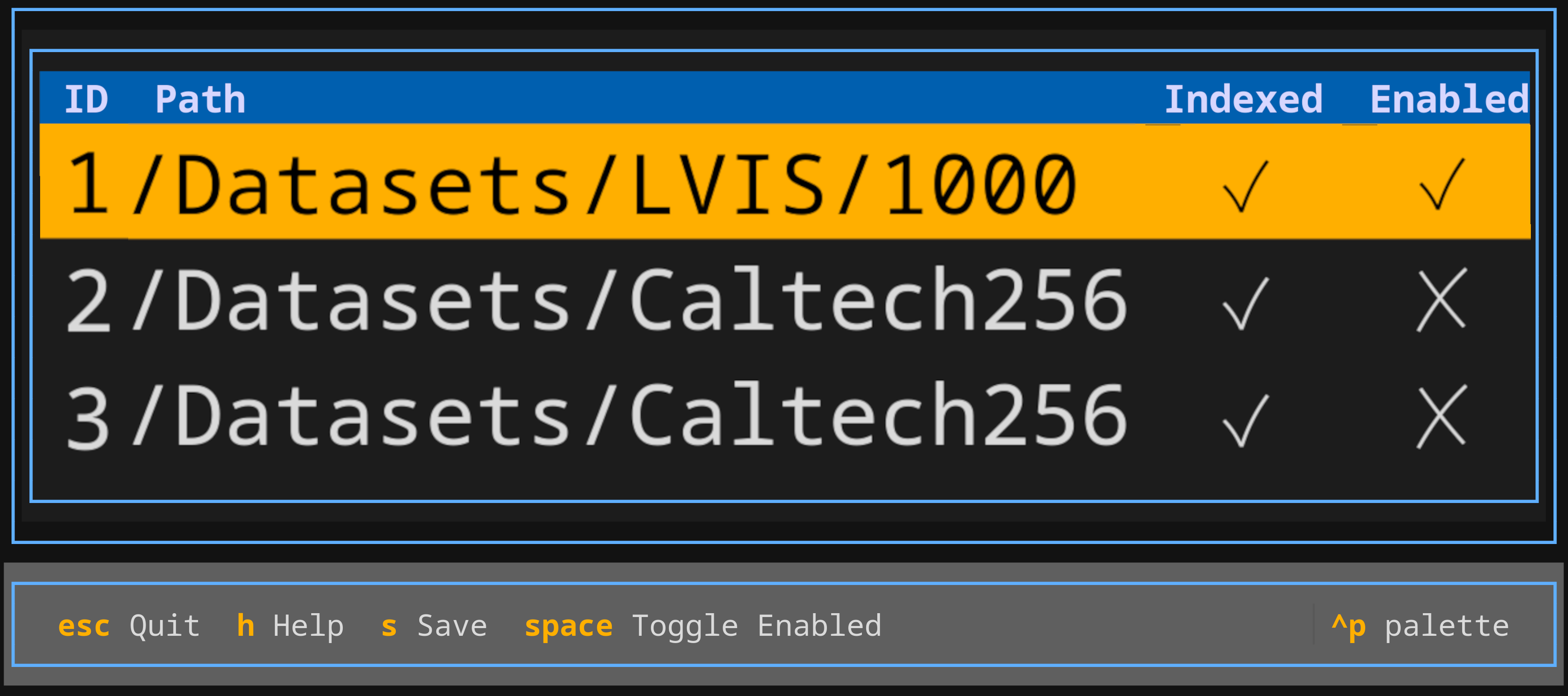}
    \caption{Directory TUI.}
    \label{fig:cli_directories}
  \end{subfigure}
  \hfill
  \begin{subfigure}{0.24\textwidth}
    \centering
    \includegraphics[height=1.95cm, keepaspectratio]{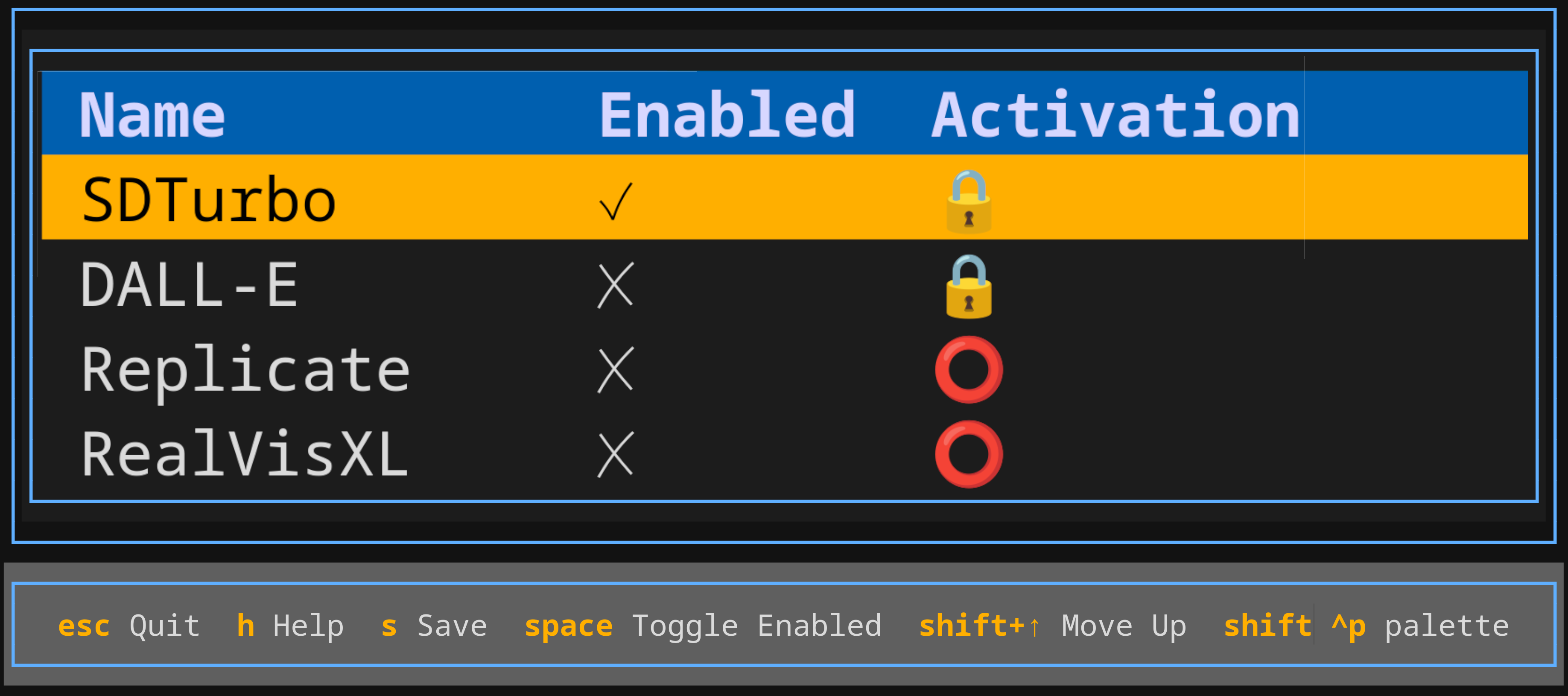}
    \caption{Generator TUI.}
    \label{fig:cli_generators}
  \end{subfigure}
  \caption{The \cli command-line interface: (a)~query execution with per-embedder verbose results, (b)~service status showing Docker containers and Python microservices, (c)~Textual-based TUI for directory management, and (d)~TUI for generator priority and activation.}
  \label{fig:cli_overall}
  \vspace{-3mm}
\end{figure*}

\noindent\textbf{Command-Line Interface (\cli).}
\cli is a single-file binary built with PyInstaller, distributed via GitHub Releases for Linux and macOS, so users need no Python environment to operate the system.
It provides five command groups spanning the full lifecycle (Figure~\ref{fig:cli_overall}):

\smallskip\noindent\emph{Service management.}
\texttt{[needlectl service start\,|\,stop\,|\,restart} {\tt |\,status]} orchestrates all infrastructure containers and Python microservices.
\texttt{[needlectl service log]} streams logs from the backend, generator hub, or Docker infrastructure.\\
\texttt{[needlectl service update]} performs in-place upgrades of individual components (the CLI binary itself, the backend, or the Web~UI) by downloading the latest GitHub release artifacts, enabling zero-downtime updates without reinstallation.

\smallskip\noindent\emph{Directory management.}
\texttt{[needlectl directory add <path>]} registers an image directory and optionally shows a live indexing progress bar.
\texttt{[needlectl directory list\,|\,describe\,|\,modify} {\tt|\,remove]} provide CRUD operations over directories.

\smallskip\noindent\emph{Query execution.}
\texttt{needlectl query run "a dog on a surfboard" -n 10} issues a search query and returns ranked results with a preview URL (Figure~\ref{fig:cli_search}); verbose output shows per-embedder rankings.

\smallskip\noindent\emph{Generator and UI management.}
\texttt{[needlectl generator list\,|\,config]} display and configure the text-to-image engines.
\texttt{[needlectl ui start\,|\,stop\,|\,build]} manage the React-based Web~UI.

\smallskip\noindent\emph{Terminal UI (TUI).}
Several \cli subcommands (\texttt{[service config]}, \texttt{[directory modify]}, \texttt{[generator config]}) launch interactive Textual-based terminal editors (Figure~\ref{fig:cli_directories}, \ref{fig:cli_generators}) where users toggle boolean flags, edit environment variables, reorder generators by priority, and save changes---all without leaving the terminal.

\smallskip\noindent\textbf{Versioning and Upgradability.}
\sys uses Git-tag-based semantic versioning: each push to \texttt{main} triggers a CI/CD pipeline that computes the version from the latest \texttt{[needlectl/v*]} tag plus the number of commits since, then builds and publishes platform-specific binaries.
Running \texttt{[needlectl --version]} reports the version of each component (CLI, backend, UI).
The \texttt{[needlectl service update]} command selectively upgrades any component by fetching the corresponding release artifact, making it straightforward to roll out patches or new embedder models without a full reinstallation.

\smallskip\noindent\textbf{Parallel Indexing.}
The backend's \emph{IndexQueueManager} uses a priority queue backed by a configurable thread pool (default: 4 workers).
Directories are processed in parallel, and within each directory, images are batched (default: 50 images per batch) so that each embedder performs a single GPU forward pass per batch.
A \emph{FileWatcherService} based on the \texttt{watchdog} library monitors registered directories in real time: newly added images are automatically queued for indexing, deleted images are removed from both PostgreSQL and Milvus, and modified images are re-embedded.
A periodic \emph{ConsistencyChecker} reconciles the database and filesystem states to handle changes that occur while the system is offline.

\smallskip\noindent\textbf{Easy Integration of New Embedders.}
Adding a new vision embedder requires only appending an entry to the \texttt{embedders.json} configuration file (specifying the \texttt{timm} model name, embedding dimension, and aggregation weight).
On the next service restart, the system automatically creates the corresponding Milvus collection and loads the model; no code changes are required.

\smallskip\noindent\textbf{Web~UI.}
A browser-based React interface (Figure~\ref{fig:overall_webui}) offers:
(a)~a \emph{search page} (Figure~\ref{fig:search}) where users type natural language queries and configure generation parameters (number of guide images, engines, image resolution), with results displayed alongside the generated guide images;
(b)~a \emph{directory management page} (Figure~\ref{fig:directories}) with live indexing progress bars, enable/disable toggles, auto-refresh, and per-directory image counts;
(c)~a \emph{generator configuration page} (Figure~\ref{fig:generators}) to enable, reorder via drag-and-drop, and parameterize text-to-image engines (including priority-based fallback); and
(d)~a \emph{status dashboard} (Figure~\ref{fig:status}) showing API health, service status, directory summaries, and generator availability at a glance.

\smallskip\noindent\textbf{Modular Image Generation.}
The \emph{Generator Hub} microservice supports six text-to-image engines out of the box---DALL-E~\cite{betker2023improving}, Replicate, RealVisXL, Imagen3-fast, local SD-Turbo~\cite{sauer2024adversarial}, and a generic \texttt{Local} HTTP endpoint---with a priority-queue-based fallback mechanism: if the highest-priority engine fails, the request is transparently rerouted to the next available engine.

\smallskip\noindent\textbf{Configuration Modes.}
The user selects one of three pre-tuned profiles at installation time, which determines the set of vision embedders, generation parameters, and index settings for the entire deployment:

\begin{itemize}[leftmargin=*, nosep]
  \item \textbf{Fast} --- 2 embedders (EVA, RegNet), 1 guide image at $512{\times}512$; optimized for low-latency retrieval.
  \item \textbf{Balanced} --- 4 embedders, 2 guide images; a middle ground between speed and accuracy.
  \item \textbf{Accurate} --- 6 embedders with learned weights, 2 guide images; highest retrieval quality.
\end{itemize}

\noindent Since the choice of embedders determines which Milvus collections are created and which models are loaded, the mode is fixed at installation time to ensure a consistent index.
The Fast configuration achieves a total inference time of \textbf{0.203\,s} per query on a 100K-image dataset, competitive with CLIP's 0.184\,s while delivering substantially higher accuracy~\cite{needle-full}.

\vspace{-2mm}
\section{Demonstration Scenarios}\label{sec:demo}

We design three interactive scenarios that highlight \sys's capabilities.
A laptop pre-loaded with the LVIS dataset~\cite{gupta2019lvis} (${\sim}$120K images, 1000+ categories) will serve as the demonstration platform.

\vspace{-2mm}
\subsection{Scenario~1: Simple vs.\ Complex Queries}
\textbf{Goal:} Illustrate the core value proposition of \sys.

An attendee first issues a simple query such as ``\emph{a red car}'' and observes that both CLIP and \sys return satisfactory results.
The attendee then composes a complex query---for example, ``\emph{a cat wearing a tiny cowboy hat sitting on a stack of books}''---and sees CLIP fail to retrieve any relevant image while \sys returns accurate results.
The interface displays the generated guide images alongside the retrieved results (Figure~\ref{fig:search}), making the generation-based retrieval mechanism visually transparent.
Attendees are encouraged to compose their own queries of varying complexity and observe how \sys's locally optimized SD-Turbo pipeline delivers sub-second responses even for challenging compositional queries.

\vspace{-2mm}
\subsection{Scenario~2: Index Your Own Dataset}
\textbf{Goal:} Demonstrate \sys's data ingestion workflow.

The attendee uses the Web~UI to register a new image directory (either a pre-loaded collection or one brought on a USB drive).
The directory management page (Figure~\ref{fig:directories}) displays a live progress bar as \sys scans, embeds, and indexes the images.
Once indexing completes, the attendee immediately runs queries against the newly indexed dataset.
We also show a side-by-side comparison of results from two pre-configured \sys instances---one in \emph{Fast} mode (2~embedders) and one in \emph{Accurate} mode (6~embedders)---to illustrate the accuracy--latency trade-off that the user commits to at installation time.

\vspace{-2mm}
\subsection{Scenario~3: Under the Hood}
\textbf{Goal:} Enable technically inclined attendees to explore \sys's configurability.

\begin{figure*}[t]
  \centering
  \begin{subfigure}{0.24\textwidth}
    \centering
    \includegraphics[height=3cm, keepaspectratio]{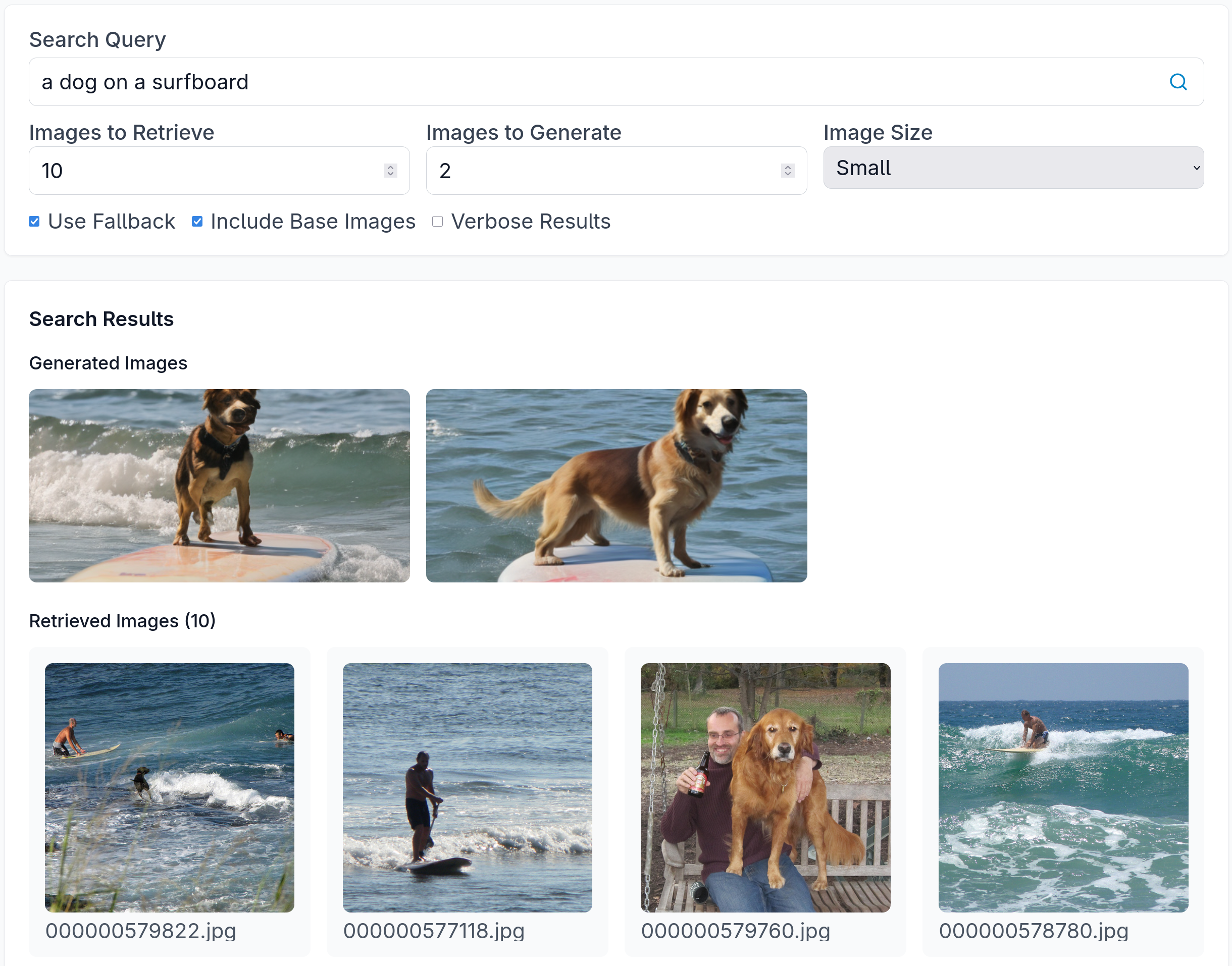}
    \caption{Search Interface.}
    \label{fig:search}
  \end{subfigure}
  \hfill
  \begin{subfigure}{0.24\textwidth}
    \centering
    \includegraphics[height=3cm, keepaspectratio]{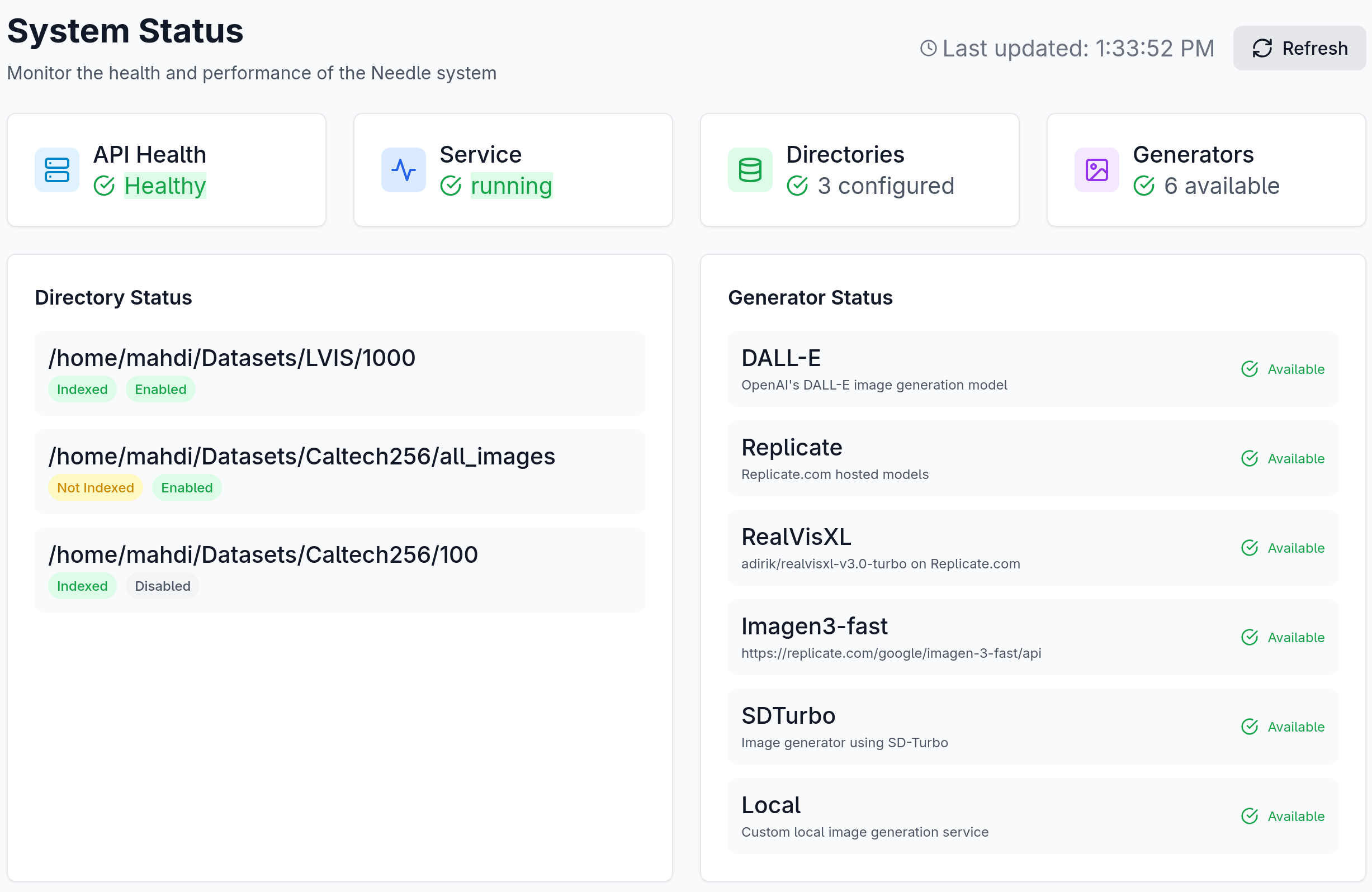}
    \caption{System Status.}
    \label{fig:status}
  \end{subfigure}
  \hfill
  \begin{subfigure}{0.24\textwidth}
    \centering
    \includegraphics[height=2cm, keepaspectratio]{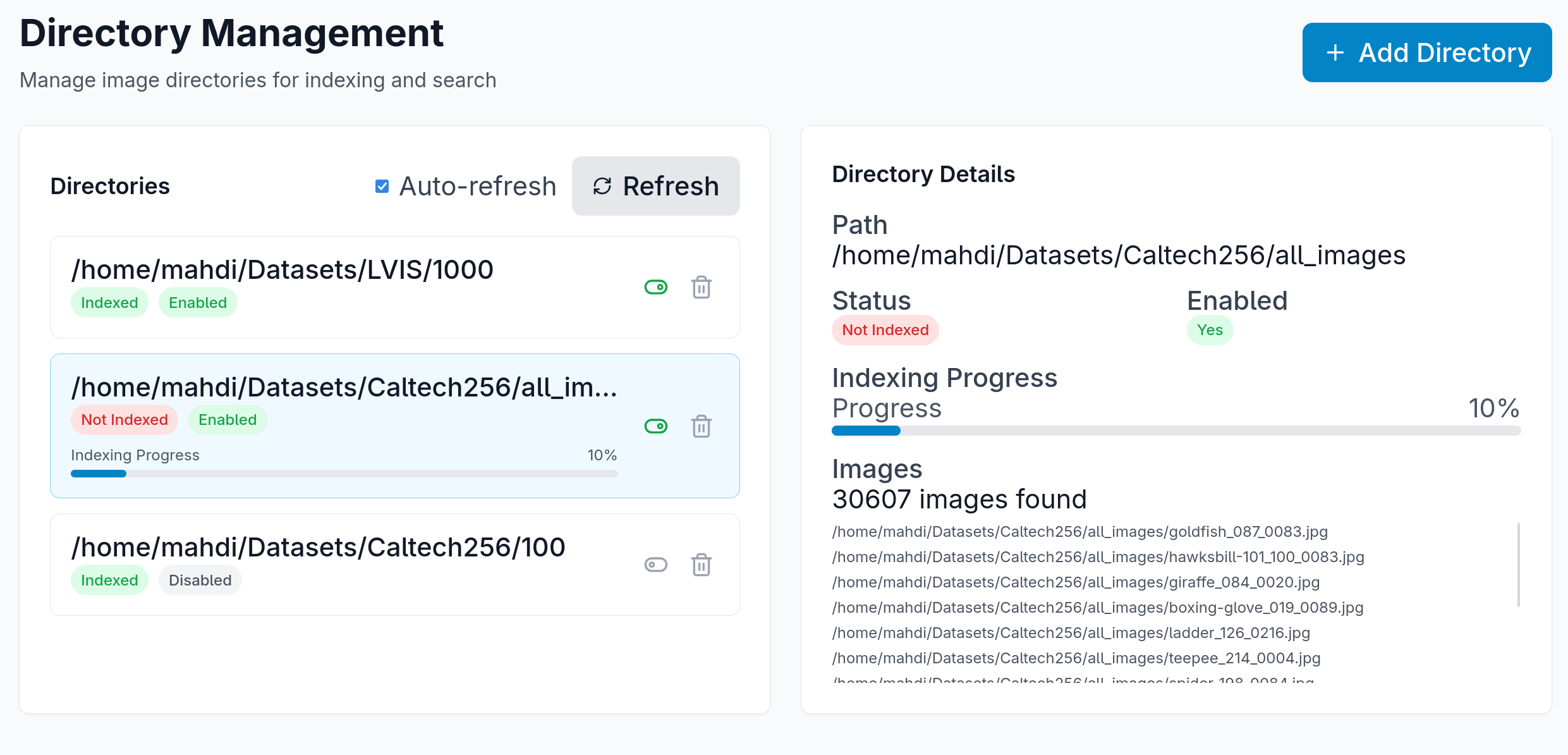}
    \caption{Directory Mgmt.}
    \label{fig:directories}
  \end{subfigure}
  \hfill
  \begin{subfigure}{0.24\textwidth}
    \centering
    \includegraphics[height=3cm, keepaspectratio]{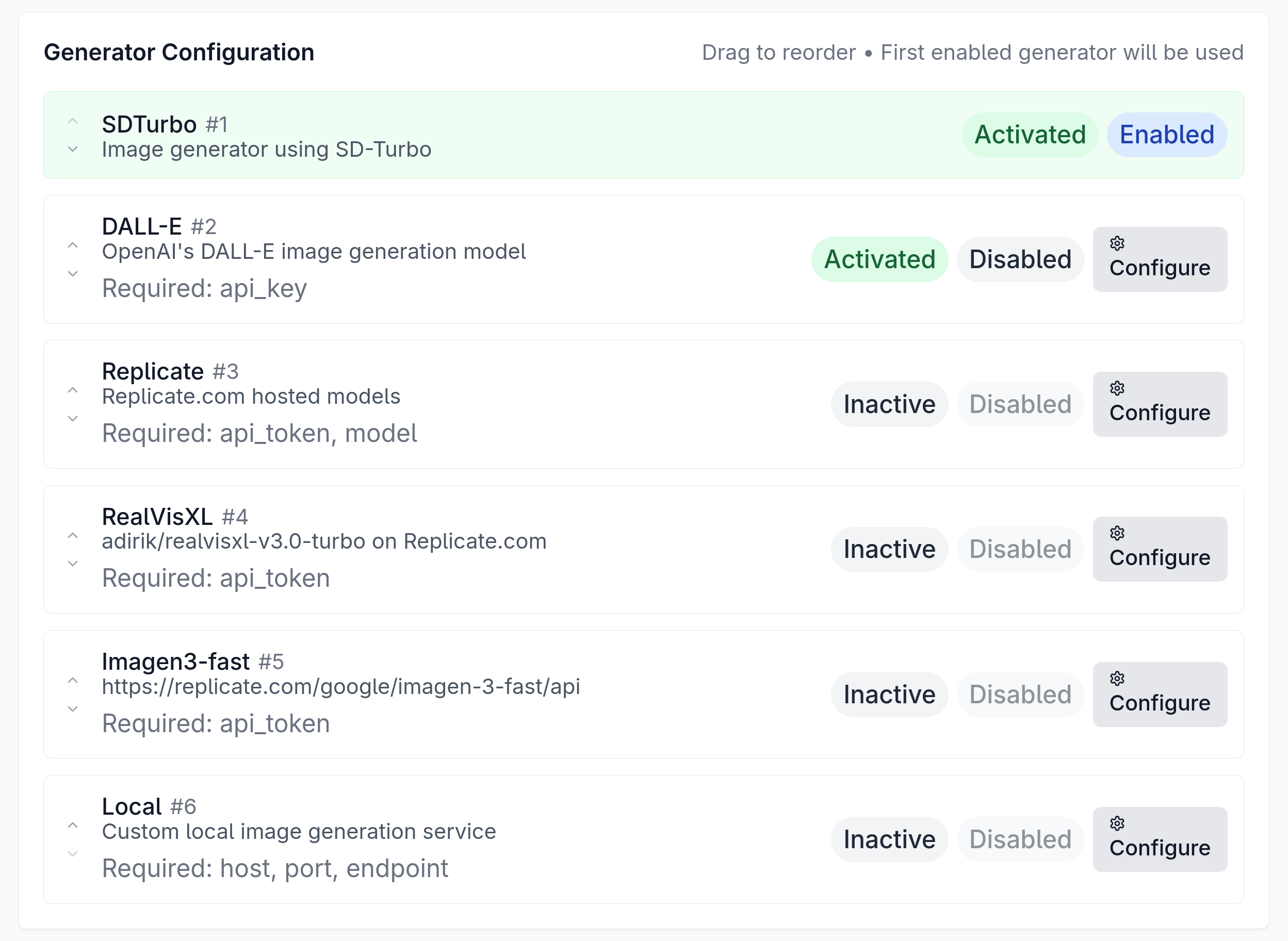}
    \caption{Generator Config.}
    \label{fig:generators}
  \end{subfigure}
  \caption{The \sys Web~UI: (a)~search page showing generated guide images alongside retrieved results, (b)~system status dashboard with API health and generator availability, (c)~directory management with live indexing progress, and (d)~generator configuration with drag-to-reorder priority.}
  \label{fig:overall_webui}
\end{figure*}

Using the CLI, the attendee varies the number of guide images ($m \in \{1, 3, 9\}$) and the image generation quality (\texttt{SMALL}, \texttt{MEDIUM}, \texttt{LARGE}).
The results panel shows individual per-embedder rankings alongside the aggregated output, making the rank-fusion process transparent.
We also compare results from different foundation models (local SD-Turbo vs.\ API-based DALL-E~3; Figure~\ref{fig:generators}) and demonstrate how the ensemble of generators mitigates individual model biases---for instance, the ``bag-of-words'' bias that ignores compositional ordering.

\vspace{-2mm}
\section{Experimental Highlights}\label{sec:results}

We summarize key results from the full experimental evaluation presented in~\cite{needle-full}.

\smallskip\noindent\textbf{Object Detection.}
Table~\ref{tab:results} reports MAP on four benchmarks.
\sys outperforms all eight baselines, with particularly large gains on \emph{hard} queries (those with CLIP AP~$< 0.5$): +73\% on Caltech256 and +93\% on LVIS relative to the second-best baseline.

\begin{table}[t]
  \caption{MAP on object detection benchmarks (All / Hard).}
  \label{tab:results}
  \vspace{-3mm}
  \centering\small
  \begin{tabular}{lcccc}
    \toprule
    \textbf{Method} & \textbf{Caltech} & \textbf{COCO} & \textbf{LVIS} & \textbf{BDD} \\
    \midrule
    CLIP~\cite{radford2021learning}     & .939\,/\,.181 & .952\,/\,.477 & .168\,/\,.078 & .670\,/\,.005 \\
    SigLIP~\cite{zhai2023sigmoid} & .965\,/\,.572 & .949\,/\,.921 & .311\,/\,.213 & .659\,/\,.067 \\
    E5-V~\cite{jiang2024e5}      & .923\,/\,.359 & .967\,/\,.951 & .327\,/\,.223 & .725\,/\,.125 \\
    \midrule
    \sys                  & \textbf{.966\,/\,.687} & \textbf{.977\,/\,.981} & \textbf{.323\,/\,.249} & \textbf{.711\,/\,.158} \\
    \bottomrule
  \end{tabular}
  \vspace{-3mm}
\end{table}

\smallskip\noindent\textbf{Complex Natural Language Queries.}
On the COLA~\cite{ray2023cola}, Winoground~\cite{thrush2022winoground}, SentiCap~\cite{mathews2016senticap}, and NoCaps~\cite{agrawal2019nocaps} benchmarks, \sys achieves the highest Pairing Accuracy (0.631 on COLA, 0.593 on Winoground) and MRR (0.642 on SentiCap, 0.745 on NoCaps), outperforming all contrastive-learning baselines~\cite{needle-full}.

\smallskip\noindent\textbf{Latency.}
The optimized Fast pipeline (SD-Turbo + EVA + RegNet) processes a query in 0.203\,s end-to-end on the LVIS dataset (100K images), of which 0.136\,s is guide image generation and 0.053\,s is multi-embedder search and retrieval.

\smallskip\noindent\textbf{Contribution Decomposition.}
An ablation study isolates the sources of improvement: switching from CLIP-based text-to-image to \sys's generative framework (using CLIP as the sole embedder) yields a 35\% relative MAP gain on LVIS; replacing CLIP with a stronger image-only embedder (EVA) adds a further 20\%; and the full ensemble provides additional incremental gains~\cite{needle-full}.

\vspace{-3mm}
\section{Conclusion}
\vspace{-1mm}
We demonstrated \sys, an open-source database system that uses generative AI to answer complex natural language queries over image datasets.
By transforming text-to-image retrieval into image-to-image search and aggregating across multiple embedders, \sys substantially outperforms existing contrastive-learning approaches while maintaining practical query latencies.
Future work includes extending the framework to audio/video and improving foundation model alignment for compositional queries.


\vspace{-4mm}
\bibliographystyle{ACM-Reference-Format}
\bibliography{ref}

\end{document}